\documentclass[11pt]{iopart}

\usepackage{iopams}  
\usepackage{hyperref}
\usepackage{graphicx}
\usepackage{epsfig}
\usepackage{url}
\usepackage{verbatim}
\usepackage{cite}

\usepackage[T1]{fontenc}

\usepackage[perpage]{footmisc}

\expandafter\let\csname equation*\endcsname\relax
\expandafter\let\csname endequation*\endcsname\relax
\usepackage{amsmath}

\begin{document}

\title[]{A percolation model for the emergence of the Bitcoin Lightning Network}
\author{Silvia Bartolucci$^1$, Fabio Caccioli$^2$, and Pierpaolo Vivo$^3$}
\address{$^1$ Department of Finance, Imperial College London Business School\\
South Kensington SW7 2AZ London (UK)\\
Centre for Blockchain Technologies, University College London\\
\vspace{10pt}
$^2$ Department of Computer Science, University College London\\
66-72 Gower Street WC1E 6EA London (UK)\\
Centre for Blockchain Technologies, University College London\\
Systemic Risk Centre, London School of Economics and Political Sciences, Houghton Street, London WC2A 2AE (UK)\\
\vspace{10pt}
$^3$ Department of Mathematics, King's College London\\
Strand WC2R 2LS London (UK)}
\ead{s.bartolucci@imperial.ac.uk, f.caccioli@ucl.ac.uk, pierpaolo.vivo@kcl.ac.uk}

\begin{abstract}
The Lightning Network is a so-called second-layer technology built on top of the Bitcoin blockchain to provide ``off-chain" fast payment channels between users, which means that not all transactions are settled and stored on the main blockchain. In this paper, we model the emergence of the Lightning Network as a (bond) percolation process and we explore how the distributional properties of the volume and size of transactions per user may impact its feasibility. The agents are all able to reciprocally transfer Bitcoins using the main blockchain and also -- if economically convenient -- to open a channel on the Lightning Network and transact ``off chain". We base our approach on fitness-dependent network models: as in real life, a Lightning channel is opened with a probability that depends on the ``fitness" of the concurring nodes, which in turn depends on wealth and volume of transactions. The emergence of a connected component is studied numerically and analytically as a function of the parameters, and the phase transition separating regions in the phase space where the Lightning Network is sustainable or not is elucidated. We characterize the phase diagram determining the minimal volume of transactions that would make the Lightning Network sustainable for a given level of fees or, alternatively, the maximal cost the Lightning ecosystem may impose for a given average volume of transactions. The model includes parameters that could be in principle estimated from publicly available data once the evolution of the Lighting Network will have reached a stationary operable state, and is fairly robust against different choices of the distributions of parameters and fitness kernels.\\
\vspace{2pt}
             
{\bf Keywords:} \noindent{\it 
Blockchain,
Lightning Network,
Payment Networks,
Percolation,
Fitness Models}
\end{abstract}

\maketitle

\section{Introduction}

	Bitcoin, the pioneering cryptocurrency, has brought about an unprecedented revolution in the payment industry \cite{satoshi}. Despite its traction and success over the last ten years, the original blockchain --  the technological infrastructure underlying Bitcoin -- suffers from some limitations that may hinder the future growth and adoption of the cryptocurrency. One of the major issue is the {\em scalability} of the system: the current number of transactions validated via this platform is between 3 and 7 transactions per second, compared for instance to thousands of transactions handled by the Visa circuit \cite{scalabilitydecker}. The lack of scalability is mainly caused by constraints on throughput of transactions, with the block size fixed at $1$MB, and by the high latency -- with a new block created on average only every ten minutes. Those limitations are imposed to safeguard the security of the platform against malicious attacks and are difficult to relax without major changes in the protocol.

	The main solutions proposed to address the scalability issue include (i) changes to the main protocol (consensus algorithm, parameters) and (ii) {\em sidechains}\footnote{Sidechains are blockchains ``connected" to the main Bitcoin blockchain such that Bitcoins can be transferred bidirectionally between the main and side blockchain \cite{franco}. At the same time, sidechains are completely separate ecosystems whose technical features or issues would not be shared with the main blockchain.} and second-layer solutions (see \cite{review} for a recent technical review). Notable examples of type-(i) solutions include new consensus protocols, which would allow a faster issuance of new blocks among other new features \cite{reviewconsensus}.  The Lightning Network (LN), instead, is a so-called second-layer technology built on top of the Bitcoin blockchain to provide ``off-chain" fast payment channels between users \cite{LNwhitepaper}. By off-chain we mean that not all transactions are settled and stored on the main blockchain.
	In a nutshell, the idea of a Lightning channel is the following: two parties lock the same amount of money as collateral and open a channel for a certain period of time. During this time, they can then exchange money back and forth through the channel, and only the netted transaction will be eventually validated and stored on the main blockchain. If one party is malicious and does not correctly update the balance, the other can keep the collateral posted by the malicious party, as a form of insurance. Any two users can open a channel and all other participants can use one or more existing channels to route transactions off-chain upon payment of a fee to channel ``owners". The scalability problem could be solved if a sufficient number of channels were opened, implying that the Lightning Network spans across the whole pool of users of the main blockchain.

	The Lightning Network topology is, indeed, relevant to understand the resilience of the system to attacks or random failures and its robustness. Measures of the network structure based on empirical data -- such as degree distribution, assortativity, shortest paths length -- provide an indication of the efficiency of payments' routing and features of the system (i.e. average number of channel per user, clusters and communities, etc.) \cite{topology1}. Experiments on random or targeted nodes removal from the network give information on the system resilience by monitoring when the original network is broken into multiple isolated clusters \cite{barrat, barabasi, internetattacks}. In the Lightning Network case, it has been shown that some types of  targeted attacks -- aimed at consuming, for instance, the channels' liquidity of specific nodes -- may yield severe consequences for the resilience of the network in terms of average payment flow and reachability \cite{topology2}. 

 	The topology of the network is in turn driven by users' economic incentives to relay transactions ``off-chain". Moreover, as the Lightning fees are set by channels' owners, an important question is how high such fees should be set in order to guarantee profits but  providing at the same time the right incentives for Bitcoin users to participate in the Lightning Network. In a recent work on simple network topologies (i.e. bidirectional channels and star graphs), the authors have estimated the demand for transactions on the main Bitcoin blockchain compared to the LN, the level of LN fees that would cover maintenance costs of the channel and their implication for the overall network security \cite{LNfees}. Indeed, transacting on the Lightning Network might impact the security of the main blockchain network by inducing a decrease in the amount of fees collected by the miners for the validation of blockchain transactions. Moreover, LN's transaction fees have been empirically studied using a traffic simulator \cite{LNsimulator}.
 
 	Fees on the main blockchain are used as incentives to miners (i.e. nodes capable of validating transactions and generating new blocks) to contribute to the security of the platform\footnote{The blockchain security is associated with the platform's decentralization, hence to the miners' total computing power \cite{mastering,MIT}.}. Normally,  users ``compete" to set up the minimal fees that would ensure that their transaction be validated within a given timeframe, as miners try to maximise the total amount of fees per block. A strand of the literature has been investigating the Bitcoin fee set-up mechanisms, the miners' incentives and their potential correlation with risks of attacks and manipulation of the transaction history.  In \cite{BTCfeesohara}  the authors use a game-theoretic model to investigate the factors influencing the value of Bitcoin fees, while in \cite{Houy} they also examine the interplay between fees and security of the platform, theoretically showing that the current fee model may not be sustainable on the long run. Alternative fee mechanisms have also been proposed, for instance based on auction models \cite{BTCfeesohara}, and compared with the existing one to highlight weaknesses and possible improvements.
 
 	The Bitcoin ecosystem has been already extensively investigated using approaches based on complex networks. The transactions network has been studied to understand latency issues and propagation mechanisms in peer-to-peer systems \cite{transpropagation} and inefficiencies of the process of permanent inclusion of the transactions on the blockchain \cite{dimatteo} .  Global and local structural properties of the users' network in Bitcoin have also provided insights on booms and bust events \cite{bubblesnetwork}
	and Bitcoin price dynamics \cite{pricenetwork}. Moreover, data on users' behaviour and spending patterns have been used to understand the global state of the crypto-economy \cite{btcnetwork4} and the drivers of the growth of the network \cite{transnetwork}. More generally, our paper taps into the growing literature on quantitative investigations of the cryptocurrencies landscape, including models of pricing and adoption of tokens \cite{econBTC,econBTC2,SBAK,baroprediction}, analysis of the market structure \cite{drozdz1,drozdz2,scirepclustering,urquhart2016, baronmachine, baronevo,marketsim} and price prediction based on sentiment and social interactions \cite{sentimentAste, tweets,sentimentplos,sentimentfrontiers,ortu,scirepwikibtc,garcia2014digital,chensentiment,yelowitz2015characteristics}.

	In this paper, we investigate under which conditions in terms of blockchain and Lightning fees, average wealth and volume of transactions per users, a Lightning Network that spans a sizeable fraction of Bitcoin users -- thus solving the scalability problem -- emerges. We model the emergence of the Lightning Network as a (bond) percolation process on a graph, exploring how different conditions may impact its feasibility  \cite{callaway}. In particular, we consider fitness-dependent network models \cite{rodgers, caldarellifit,butta, bianconi} where the probability of creating a new edge depends on intrinsic node features collectively denoted {\em node fitness}. In the LN case, the node fitness will be defined in terms of the node wealth and activity (i.e. volume of transactions).
 	The viability of the Lightning Network will be characterized in terms of the presence (or not) of a giant connected cluster of nodes: a non-fragmented network would, indeed, guarantee a smooth relay of payments and information between users and will incentivize off-chain transactions. Our model depends on parameters that can be all obtained -- or at least estimated -- from publicly available data, and is fairly robust against different choices of distributions of parameters and fitness kernels.

The paper is organized as follows. In Section \ref{sec:blockchain} we provide a quick overview of the main Bitcoin blockchain and the main ideas behind LN. In Section \ref{sec:model} we describe our model and provide the relevant theory, which is then applied to two specific wealth distributions (uniform and exponential) in the subsections \ref{sec:uniform} and \ref{sec:exp}, respectively. In Section \ref{sec:simresults}, we discuss the results of numerical simulations, and we provide some conclusions and outlook in Section \ref{sec:conclusions}. The Appendices are devoted to technical aspects of percolation theory on networks and are included to make the paper self-contained.

\section{The Bitcoin blockchain and Lightning Network} \label{sec:blockchain}

	In this section, we summarize the main features of the Bitcoin main blockchain and Lightning Network payment layer.
	The Bitcoin blockchain is a distributed, shared ledger that immutably records transactions among peers in the network \cite{mastering, satoshi}. Transactions are bundled in blocks and chained together via cryptographic primitives to ensure that any change at any point in the transaction history would invalidate the full record. Transactions are validated for correctness, temporarily stored in memory pools and then arranged in the blocks data structure by \emph{miners}: multiple miners compete using computational power to validate the next block of the chain -- and therefore earn the associated reward for the service and transactions' fees--according to the Proof-of-Work consensus algorithm. 
	Depending on the usage of the network and due to limitation in block size, waiting times can peak around $~30$ minutes (while the typical range is around $~6-8$ minutes), while blockchain fees per transaction exhibit a broad range of variability, from a few cents to $~40-50$ USD\footnote{Data taken from \url{https://www.blockchain.com/charts} .}.

\begin{figure}[htb!]
\centering
\includegraphics[width=0.48\textwidth]{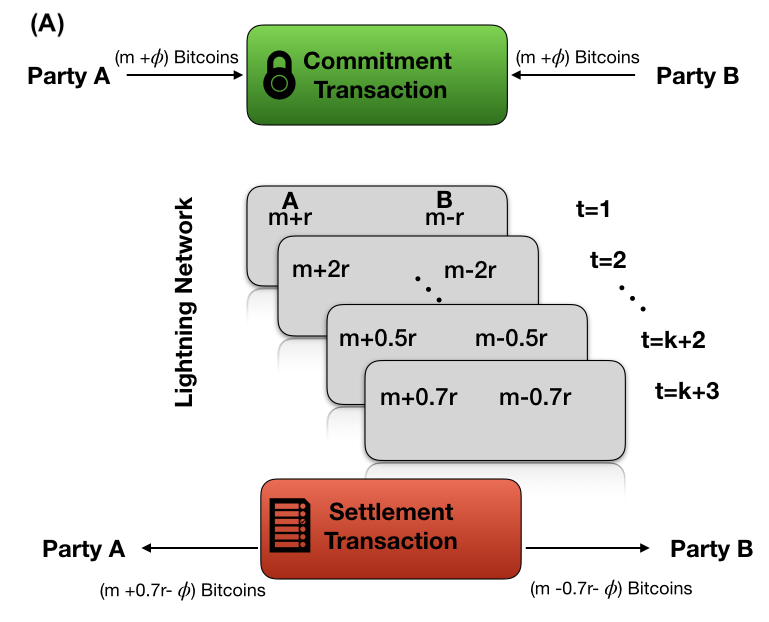}
\includegraphics[width=0.48\textwidth]{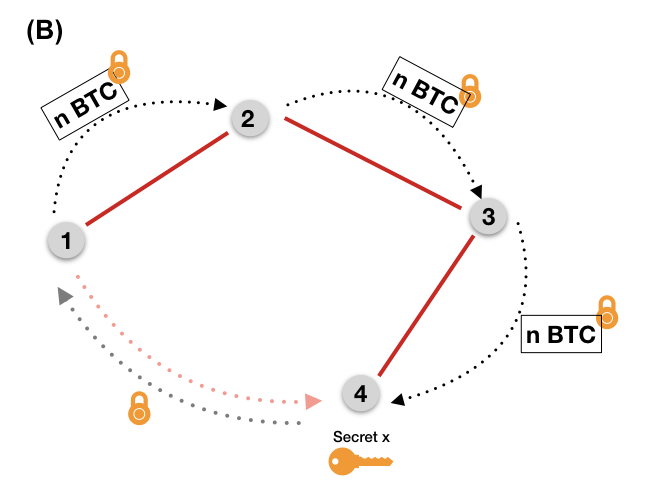}
\caption{Panel {\bf (A)}: Scheme of a payment channel between party A and B, including opening and closing transactions settled on the main blockchain, and intermediate transfers handled off-chain on the Lightning Network. Initially, the two parties lock $m$ bitcoins each, plus the fee $\phi$ that would cover broadcasting on the main blockchain. At each time step $t$, users exchange an amount $\alpha(t)r$ and update their balances accordingly, by sending each other redeemable receipts without committing them on the main blockchain. Only when the two parties agree that the channel is no longer needed, they settle the net balance of funds on their original Bitcoin addresses.  
Panel {\bf (B)}: Scheme of payments routing on a Lightning Network: even if party 1 and 4 are not directly connected via an existing LN channel, they can route their payments through other parties (upon payment of a fee) by choosing a suitable cryptographic lock for the Bitcoins.}
\label{fig:LNchannels}
\end{figure}
The idea behind the creation of the Lightning Network \cite{LNwhitepaper} is, therefore, to devise a network for frequent and fast micro-transactions that can be performed at low transactions fees. The basic components of the Lightning Network are {\em payment channels} (schematically shown in Fig. \ref{fig:LNchannels}, panel A), enabling trustless transfers between users. In the typical payment channel implementation, a theoretically unlimited amount of payments can be made, with only two transactions broadcast on the blockchain. In addition to a reduction of  number of blockchain transactions and associated costs, payment channels also offer the advantage of speed and, importantly, the ability of users to recover their funds if one of the parties is malicious. 

	A channel is established between two parties by locking an initial amount of funds, for instance $m$ Bitcoins for each user, on the main blockchain, which represent the maximum amount of Bitcoins that can be transferred over the channel. Funds are locked on so-called 2-of-2 multisignature addresses \cite{mastering}, which can be unlocked upon providing the signature of both interested parties. For instance, user A wishes to send $r$ Bitcoins to user B: she signs a transaction, sends it to B, who will sign it and send it back to A. Only the first transaction is recorded on the main blockchain. At each time step in the lifetime of the channels, the users keep sending back and forth signed transactions that can be at any point consensually broadcast on the main blockchain to close the channel and redeem the net amount of funds. To prevent fraudulent behavior, for instance user B not acknowledging the receipt of a payment from A, a refund option is always included in any exchange. The refund option can be unilaterally unlocked and submitted to the blockchain after a certain amount of time $t^{\star}$ has elapsed from the moment the channel was first established. Every new refund option is indeed signed by both parties, signaling therefore that they are in agreement with the terms of the refund, which may be exercised unilaterally at a later time. In the worst-case scenario, one party would simply submit the original refund transaction created contextually with the opening of the channel.

	Payments can be relayed via the Lightning Network also if two parties are not directly connected via a Lightning channel, if there exists a path indirectly linking them via existing channels owned by third parties. Exploiting an existing path to route the payments may often prove more convenient as the two interested parties need not open a new channel, therefore saving the associated costs in blockchain fees. Channels' owners are indeed owed  ``routing fees" to allow payments through their channel, but at the moment those fees are very competitive ($\sim 4$ orders of magnitude less than the Bitcoin blockchain\footnote{Data taken from \url{https://1ml.com/statistics} and \url{https://bitcoinfees.info} .}). In Fig. \ref{fig:LNchannels}, Panel B we show an example of an indirect routing path between user $1$ and $4$.
	One of the biggest issues of the Lightning Network is the limitation in liquidity. Payments are made by effectively having intermediaries forwarding collateral across multiple channels: this means that if party 1 is transferring $\ell$ Bitcoins to party $4$, each relaying channel needs to have at least $\ell$ Bitcoins available in the direction of the payment. 
	
	To prevent dishonest behavior in the transfer from party 1 to  4 via party 2 and 3 (see Fig. \ref{fig:LNchannels}, Panel B), Party 1 will lock the Bitcoins with a secret key known only by the receiver: when party 4 receives the Bitcoins from party 3, the secret is revealed and every player can collect their coins and fees \cite{mastering}. 
	
	\section{Model Setup} \label{sec:model}
In this section, we model the emergence of the Lightning Network as a (bond) percolation process. We consider $N$ agents, who are all able to reciprocally transfer Bitcoins using the main blockchain and -- if economically convenient -- to open a channel on the Lightning Network and transact ``off chain".
We introduce the node capacity (or \emph{wealth}) $w_{i}$ of node $i$, a random variable extracted from a pdf $\Pi(w)$, which is proportional to the maximum amount of Bitcoins that node $i$ can lock in a Lightning channel it partakes in. We will consider two explicit examples for the wealth distribution (uniform and exponential) in the following, with qualitatively similar results.

Two nodes are more likely to open a Lightning channel if they expect to submit a large number of transactions over a given period of time. Therefore, we introduce for each node $i$ a quantity $\ell_i$ that represents its ``activity" in terms of average number of transactions node $i$ sends through each channel in the network. The average number of transactions is also a random variable extracted from the discrete distribution $\hat{\Pi}(\ell)$ over non-negative integers. We also include the costs associated with transacting over one of the two networks (main blockchain only or blockchain and Lightning). These costs can be fixed per transaction ({\em base fee}) or can be calculated as a percentage of the value transferred ({\em fee rate}). 
\begin{itemize}
\item $c$, \emph{Lightning channel maintenance/usage base fee:}\\
Using the LN channel provided by an operator or  other users to transfer coins carries an associated LN fee. Opening a channel has also maintenance costs (fee setup, market and nodes monitoring, connections) and costs related to locking Bitcoins and providing liquidity in the channel.\\
\item  $\phi$, \emph{main blockchain rate fee:}\\
We assume that a fraction $\phi$ of the value transferred in each transaction needs to be paid by the sender to have it included in blocks and validated by miners. \\
\end{itemize}

The probability of opening a new LN channel between two nodes $p_{ij}^{\mathrm{LN}}$ can be modeled as a function of (i) the costs associated with opening the LN channel (if the costs are significantly smaller than using the Bitcoin blockchain, there is an incentive for the users towards opening the channel), (ii) users' affinity (the more likely are users to transact over a period of time $\tau$, the higher the benefits of opening a channel), (iii) the wealth of the nodes (nodes wishing to open a LN channel have to lock a minimal amount of Bitcoins on the main blockchain as collateral).

The growth of the Lightning Network can be modeled as a bond percolation process on a set of $N$ nodes representing Bitcoin users. The edges then represent new Lightning channels being opened.
In particular, we construct the bond percolation model considering fitness-dependent networks \cite{rodgers,caldarellifit,butta,bianconi}. In fitness models, the network topology is determined by (i) an {\em attachment kernel} $f(x,y)$, describing the probability that a node with fitness $x$ will connect to a node with fitness $y$, and (ii) the distribution of fitness $\rho(x)$ across nodes.

The network we consider has a fixed number of nodes $N$ -- corresponding to all Bitcoin users that may decide to switch to the LN -- and is \emph{sparse}, i.e. the number $M$ of edges is $M\ll N^2$. If we consider node $i$ and $j$ having fitness $x_i$ and $x_j$ respectively, a LN channel, i.e. an edge between them, is added with probability
\begin{equation}
p_{ij}^{\mathrm{LN}}= f(x_i, x_j)\sim\mathcal{O}(1/N) \ .\label{probF}
\end{equation}
The resulting network is undirected if $f(x,y)=f(y,x)$, which is a sensible requirement: indeed, opening a LN channel between two nodes will require a ``symmetric" commitment from both nodes to lock Bitcoins on the main blockchain. In our model, we will consider bond percolation only: number and ``state" of the nodes (e.g. occupied/unoccupied or infected/susceptible) will not change.

In the context of the LN network, we define the fitness $x_i$ of node $i$ as the simplest increasing function of both capacity and volume of transactions, i.e. 
\begin{equation}
x_i=w_i (\ell_i+1)\ , 
\end{equation}
where $w_i$ represents the wealth of the node, and $\ell_i+1\geq 1$ its ``activity" in terms of number of transactions expected to be sent through the channel\footnote{We assume that all nodes are potentially active, $x_i>0$ for all $i$.}. As in \cite{caldarellifit}, we consider the fitness to be defined in the interval $[0,\infty]$.

Given this definition of the node fitness, the fitness distribution can be calculated from the wealth and activity distributions, $\Pi(w)$ and $\hat{\Pi}(\ell)$ respectively, as 
\begin{equation}
\rho(x)= \sum_{\ell\geq0} \hat{\Pi}(\ell) \int_{0}^{\infty}{\rm d}w~ \Pi(w) \delta\left(x- w(\ell+1)\right) \ .
\label{eq:rho}
\end{equation}
If we imagine links are added one at a time at a given rate, from the kernel $f(x,y)$ we can derive the probability that a node with fitness $x$ increases its degree by one as \cite{rodgers}
\begin{equation}
\lambda(x,N)=\frac{1}{N}\frac{\int_0^\infty{\rm d }y f(x,y)\rho(y)}{ \kappa}\ ,
\end{equation}
where $N\kappa\sim\mathcal{O}(1)$ is the average degree of the network with $N$ nodes, with
\begin{equation}
\kappa = \int_0^\infty\int_0^\infty {\rm d }x {\rm d }y~\rho(x)\rho(y)f(x,y)\ .\label{kav}
\end{equation}
We also define $\lambda(x)= N\lambda (x,N)$ and rewrite it as
\begin{equation}
\lambda(x)=\frac{1}{\kappa}\int_0^\infty {\rm d }y~f(x,y)\rho(y)\ . \label{deflambda}
\end{equation}
Note that $\lambda(x)$ clearly satisfies the following normalization condition
\begin{equation}
\int_0^{\infty}\lambda(x)\rho(x){\rm d}x=1\ .\label{normlambda}
\end{equation}
The degree distribution $P(k)$ for large $N$ is given by
\begin{equation}
P(k) = \int_0^\infty {\rm d}x~\rho(x)\frac{{\rm e}^{-N\kappa\lambda(x)}[N\kappa\lambda(x)]^k}{k!}\ ,\label{Pk}
\end{equation}
whose average degree is $\langle k\rangle = N\kappa$, as shown in detail in \ref{app:degree}.

In the following, we will assume that the activity distribution is Poisson with average $\bar{n}$, $\hat\Pi(\ell)= \exp(-\bar n)\bar n^\ell/\ell! $, and that the connectivity kernel models the effects of blockchain and LN fees as follows
\begin{equation}
f(x,y) = \frac{\mu}{N}\Theta(x\phi-c)\Theta(y\phi-c)\ ,\label{definitionf}
\end{equation}
where $\Theta(z)$ is the Heaviside step function\footnote{We have checked that smoothing $f(x,y)$, e.g. by multiplying the thetas by $xy/(1+xy)$ or $1-\exp(-(x+y))$, has a minimal impact on the results.}. The interpretation of this kernel is as follows: agent $i$ expects to interact with $\mu$ other agents, which we assume for simplicity are chosen randomly.

\begin{figure}[htb!]
\centering
\includegraphics[width=0.44\textwidth]{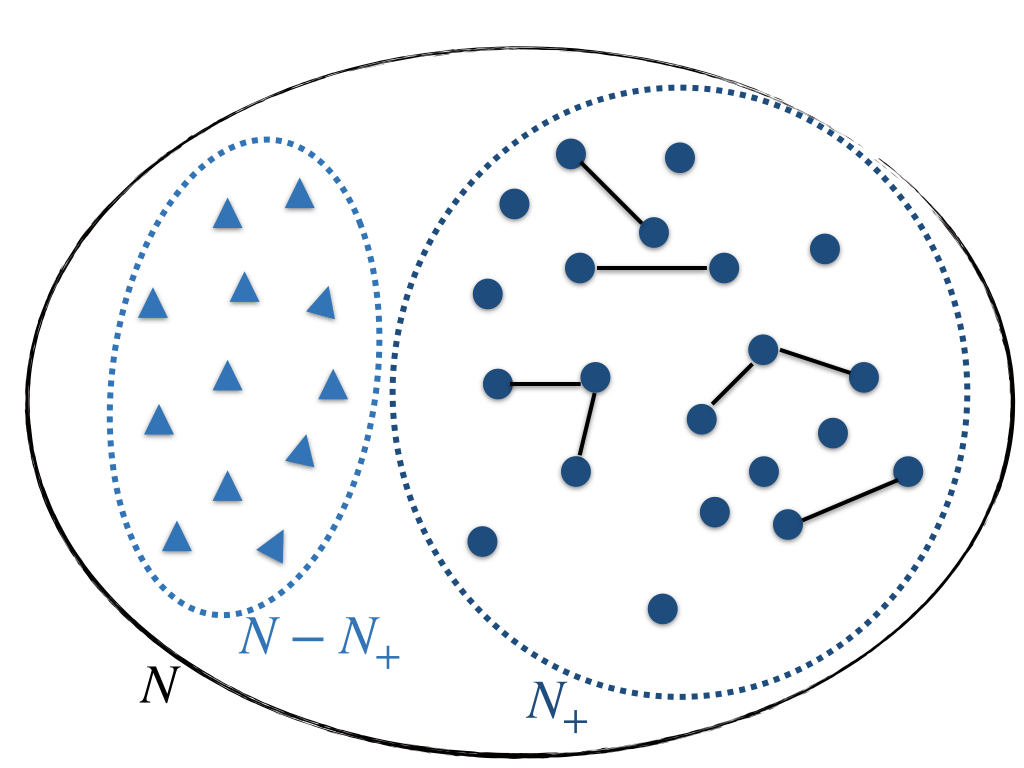}
\includegraphics[width=0.44\textwidth]{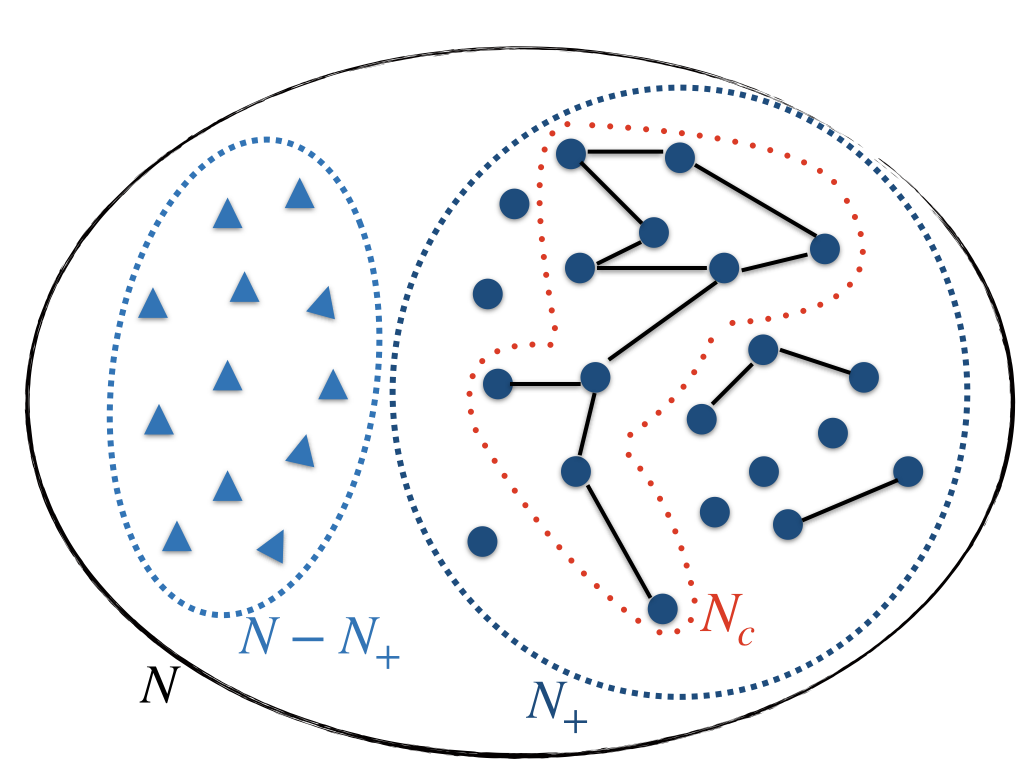}
\caption{Schematic representation of the emergence of the connected component among fit nodes, below (left) and above (right) the percolation threshold. Light blue triangles represent nodes whose fitness is smaller than $c/\phi$, while dark blue circles represent ``high-fitness" nodes with $x>c/\phi$. $N$ is the total number of Bitcoin users, $N_+$ is the fraction of ``high-fitness nodes" (see Eq. \eqref{Nplus}) and $N_C$ is the fraction of high-fitness nodes belonging to the giant connected component.}
\label{fig:schemeperc}
\end{figure}

The probability of interacting with a given agent $j$ is equal to $\mu/N$ for all $j$. Agent $i$ wishes to transfer an amount $w_i (\ell_i+1)$ (corresponding to $\ell_i +1$ transactions of size $w_i$) to each of them, and is willing to open a Lightning channel if the cost of maintaining it ($c$) is lower than the cost of transferring the money through the blockchain ($w_i (\ell_i+1)\phi)$. The same considerations apply to its counterpart $j$. 

We define
\begin{equation}
N_+ = \sum_{i=1}^N \Theta(x_i\phi-c)
\label{Nplus}
\end{equation}
the number of nodes with ``high" fitness, for whom it is economically viable to engage in a LN. Note that $N_+$ is a random variable, which depends on the realization of the fitnesses. We define the average fraction $f_+ = \langle N_+\rangle/N$.

The network constructed via the sequential deposition of links (as described above) may undergo a \emph{percolation transition}  \cite{newman,newman2,rodgers,callaway} as a function of $f_+$, such that -- beyond a critical value of $f_+$ -- a giant connected component of $N_C$ nodes emerges, whose fractional average size $S=\langle N_C\rangle/N$ remains finite as $N\to\infty$. We stress that in any fixed instance $N_C\leq N_+$, since some high-fitness nodes may still not engage in LN (see Fig. \ref{fig:schemeperc}). In our language, this connected component represents the set of nodes that not only do exploit Lightning channels to exchange wealth off-chain between nearest neighbors, but may also transfer wealth to any ``distant node", routing the transaction via connected paths. It is therefore of paramount importance to understand under which conditions on the average wealth, average volume of transactions, and routing fees, this transition may happen, and what finite fraction of nodes will it involve. 

With the choice of the kernel in \eqref{definitionf}, the topology of the resulting Lightning Network of $N_+$ nodes is that of an Erd\H{o}s-R\'enyi (E-R) graph with average degree equal to $\mu f_+\sim\mathcal{O}(1)$. 
At odds with the standard model of E-R graphs, in our case the size of the graph $N_+$ is itself a random variable, which depends on the parameters of the model. In fact, once $f_+$ has been obtained, the model can be mapped onto a site percolation problem on random networks, where each node is occupied with probability $f_+$, and the emergence of a viable Lightning Network corresponds to the emergence of a giant component of occupied nodes \cite{dorogovtsev2008critical}.

The relevant percolation theory is summarized in \ref{app:giantc} to make the paper self-contained.

\subsection{Uniform wealth distribution} \label{sec:uniform}
We now take $\Pi(w)$ -- the pdf of wealth across nodes -- as uniform in the interval $[0, w_0]$. Hence, we have
\begin{equation}
\rho^{(u)}(x)= \sum_{\ell\geq0} \frac{{\rm e}^{- \bar{n}}\bar{n}^\ell}{\ell!} \int_{0}^{w_0}\frac{{\rm d}w}{w_0} \delta\left(x- w(\ell+1)\right)\ ,
\end{equation}
where the superscript $^{(u)}$ refers to uniform wealth distribution.
Simplifying we obtain
\begin{equation}
\rho^{(u)}(x)=\frac{1}{w_0} \sum_{\ell\geq\lceil\frac{x}{w_0}-1\rceil} \frac{{\rm e}^{- \bar{n}}\bar{n}^\ell}{\ell!(\ell+1)} =\ \frac{1}{w_0 {\bar n}}\left( 1-  \frac{\Gamma \left(\lceil\frac{{x}}{w_0}\rceil,\bar{n} \right)}{\Gamma \left(\lceil\frac{{x}}{w_0}\rceil\right)}\right)\ ,
\end{equation}
where $\Gamma(a,x)=\int_x^{\infty} t^{a-1}{\rm e}^{-t}{\rm d}t$, and $\lceil z\rceil$ denotes the smallest integer larger than $z$.
In this case, it follows from \eqref{deflambda} and \eqref{definitionf} that
\begin{equation}
\lambda^{(u)}(x) = \frac{1}{f^{(u)}_+}\Theta(x\phi-c)\ ,
\end{equation}
where $f^{(u)}_+$ is the average fraction of high-fitness nodes and is given by
\begin{equation}
f^{(u)}_+ = \int_{c/\phi}^\infty {\rm d}x~\rho^{(u)}(x)=
 \frac{1}{w_0}\sum_{\ell\geq 0}\frac{{\rm e}^{-\bar n}\bar n^\ell}{\ell!}\int_0^{w_0}{\rm d}w~\Theta(w(\ell+1)-c/\phi)\ ,\label{Cudef}
\end{equation}
which requires $w>c/[(\ell+1)\phi]$, in turn constraining $c/[(\ell+1)\phi]\leq w_0\rightarrow\ell\geq \lceil\frac{c}{w_0\phi}-1\rceil$ (which may also be negative). Therefore
\begin{equation}
f^{(u)}_+ = \frac{1}{w_0}\sum_{\ell=\max\left(0, \lceil\frac{c}{w_0\phi}-1\rceil\right)}\frac{{\rm e}^{-\bar n}\bar n^\ell}{\ell!}\int_{\frac{c}{(\ell+1)\phi}}^{w_0}{\rm d}w=\Psi(0)-\frac{c}{\phi w_0}\Psi(-1)\ ,
\end{equation}
where
\begin{equation}
\Psi(t) = \sum_{\ell=\max\left(0, \lceil\frac{c}{w_0\phi}-1\rceil\right)}\frac{{\rm e}^{-\bar n}\bar n^\ell}{\ell!}(\ell+1)^t\ .
\end{equation}
The evaluation of $P^{(u)}(k)$ from \eqref{Pk} requires some care, as $\lambda^{(u)}(x)$ is zero if $x<c/\phi$. Splitting the integration region, we get
\begin{align}
\nonumber P^{(u)}(k) &=\delta_{k,0}\int_0^{c/\phi}{\rm d}x~\rho^{(u)}(x) + \frac{{\rm e}^{-\mu f^{(u)}_+}(\mu f^{(u)}_+)^k}{k!}\int_{c/\phi}^\infty {\rm d}x~\rho^{(u)}(x)\\
&=\left(1-f^{(u)}_+\right)\delta_{k,0}+f^{(u)}_+\frac{{\rm e}^{-\mu f^{(u)}_+}(\mu f^{(u)}_+)^k}{k!}\ .\label{Pkunif}
\end{align}
The interpretation of \eqref{Pkunif} is quite neat: on average, the network contains a fraction $1-f_+$ of isolated (low-fitness) nodes, and a fraction $f_+$ of high-fitness nodes that may (or may not) partake in the LN, establishing sparse random connections with an average of $\mu f_+$ other high-fitness nodes.
Computing now the generating function \eqref{G0s}
\begin{equation}
G_0^{(u)}(s)=1-f^{(u)}_+ +f^{(u)}_+\exp\left(\mu f^{(u)}_+(s-1)\right)\ ,
\end{equation}
it follows from Eq. \eqref{G1s} that
\begin{equation}
G_1^{(u)}(s) = \frac{G_0^{(u)\prime}(s)}{G_0^{(u)\prime}(1)}=\exp\left(\mu f^{(u)}_+(s-1)\right)\ .
\end{equation}
The general theory (see \ref{app:giantc}, in particular Eq. \eqref{xistar}) then implies that the equation determining $0<\xi^\star\leq 1$ is
\begin{equation}
\xi^\star = \exp\left[\mu f^{(u)}_+(\xi^\star-1)\right]\ .
\end{equation}
The average size of the giant component thus reads from Eq. \eqref{xistarsize}
\begin{equation}
S^{(u)}=(1-\xi^\star)f^{(u)}_+\ ,
\end{equation}
and the condition in Eq. \eqref{G1diverge} for the giant component to appear is
\begin{equation}
\mu f^{(u)}_+>1 \ .
\label{mufplus}
\end{equation}
The interpretation of this condition is fairly obvious: the giant connected component can only arise if ``fit" nodes open on average more than one channel with other fit nodes. 

\begin{figure}[h!]
\centering
\includegraphics[width=0.9\textwidth]{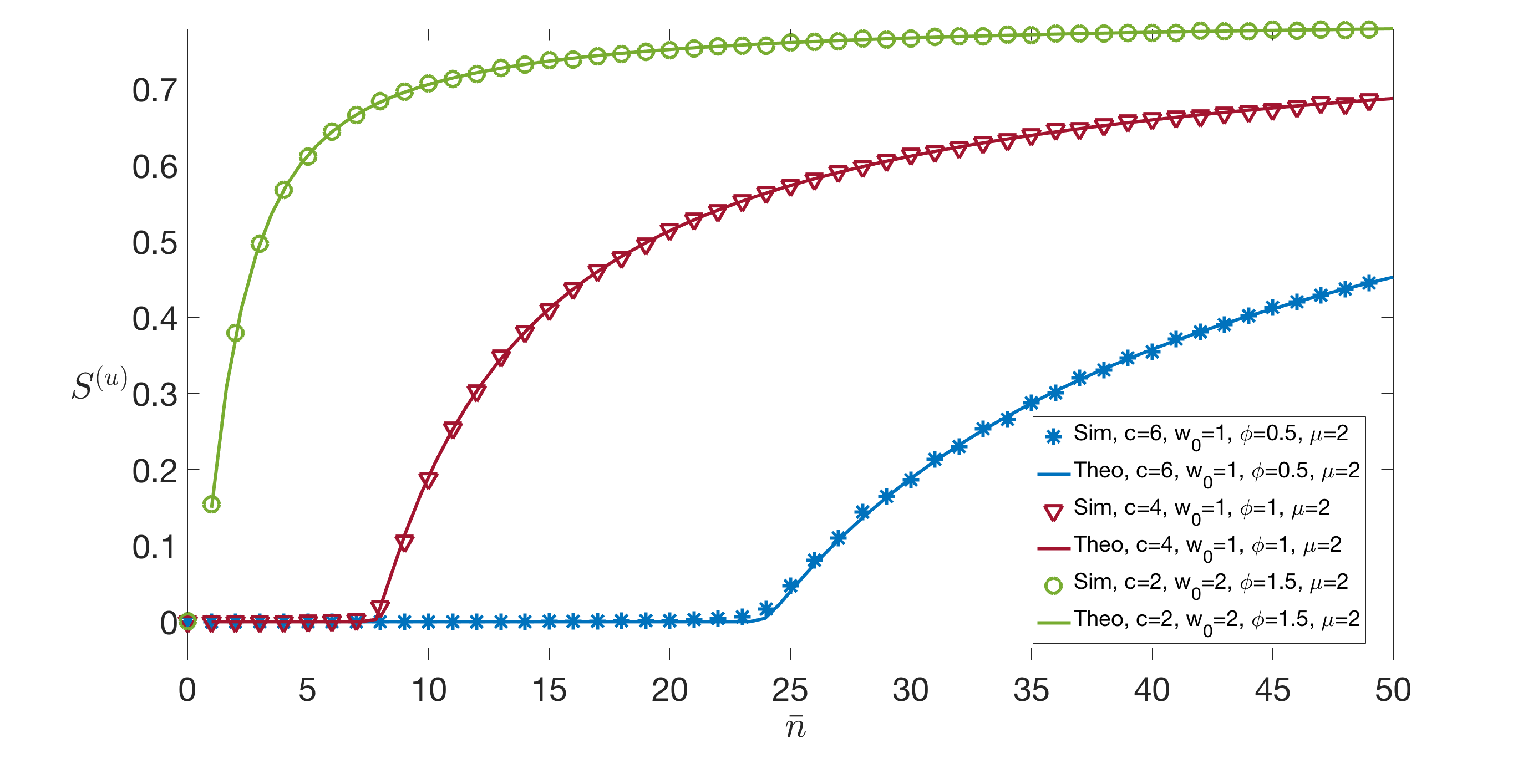}
\caption{Size of the giant component as a function of $\bar{n}$ for different combinations of the parameters $\phi, c,w_0, \mu$.  Simulations with kernel $f(x,y)$ in Eq. \eqref{definitionf} and uniform wealth distribution in the interval $[0,w_0]$. Numerical results have been obtained for a network of $N=5\cdot 10^4$ nodes averaging over $5$ independent instances. Fixing a certain fraction $S$ of nodes -- connected via LN -- and increasing the ratio $c/\phi$ between the Lightning Network and main blockchain fees, we observe that a larger average volume of LN transactions is required to make the system financially sustainable. The transition between $S=0$ and $S>0$ happens at a value of $\bar n$ that we denote $\bar n^\star$.}
\label{fig:unifsize}
\end{figure}

\begin{figure}[h!]
\centering
\includegraphics[width=0.9\textwidth]{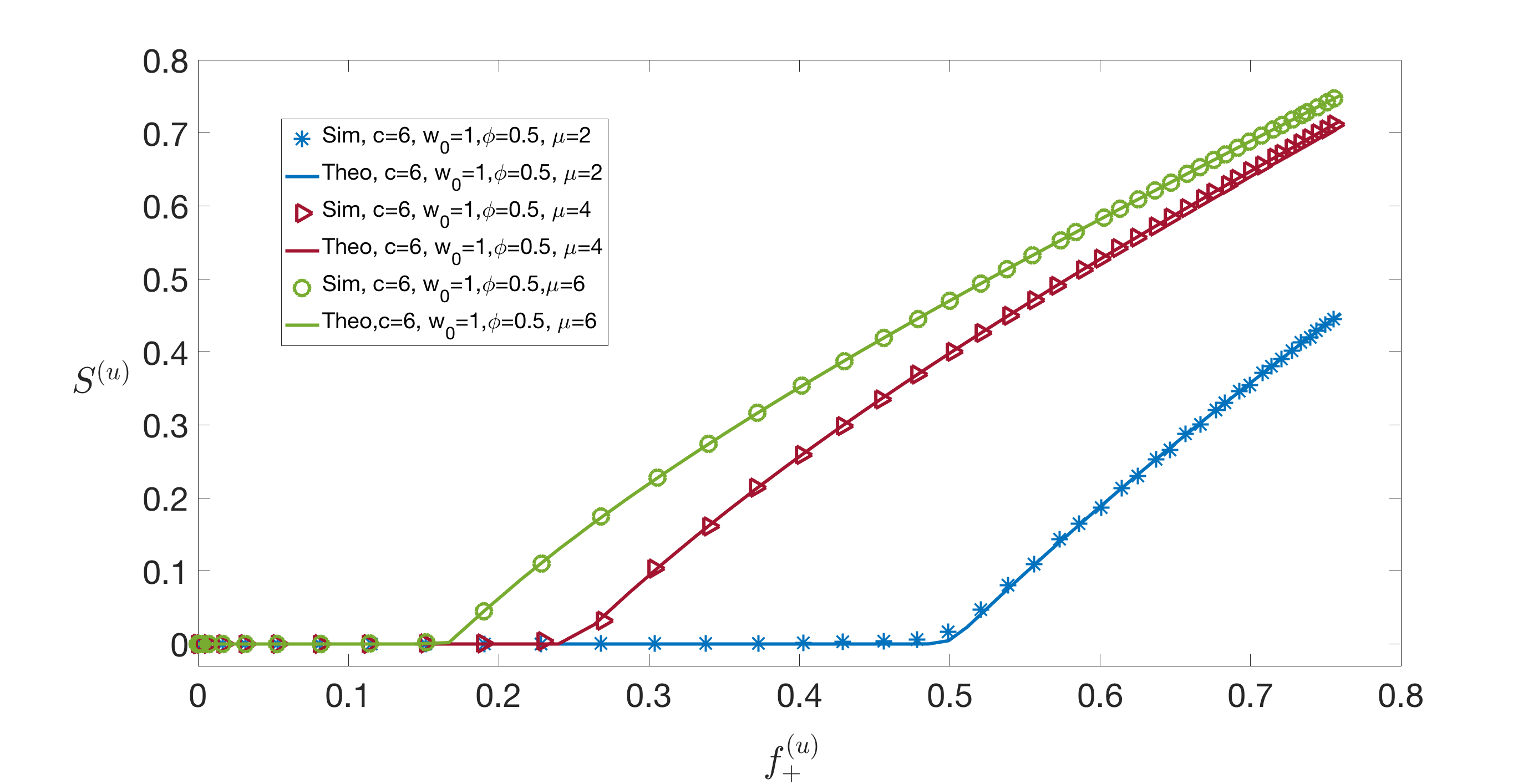}
\caption{Size of the giant component as a function of $f^{(u)}_+$ (the fraction of high-fitness nodes) varying $\mu=2,4,6$ and fixing $w_0=1,\phi=0.5, c=6$. Simulations with kernel $f(x,y)$ in Eq. \eqref{definitionf} and uniform wealth distribution with parameter $w_0$. Numerical results have been obtained for a network of $N=5\cdot 10^4$ nodes averaging over $5$ independent instances. The transition between $S=0$ and $S>0$ happens at $1/\mu$ as correctly predicted by the condition in Eq. \eqref{mufplus}.}
\label{fig:unifsize}
\end{figure}

\subsection{Exponential wealth distribution} \label{sec:exp}

\begin{figure}[h!]
\centering
\includegraphics[width=0.9\textwidth]{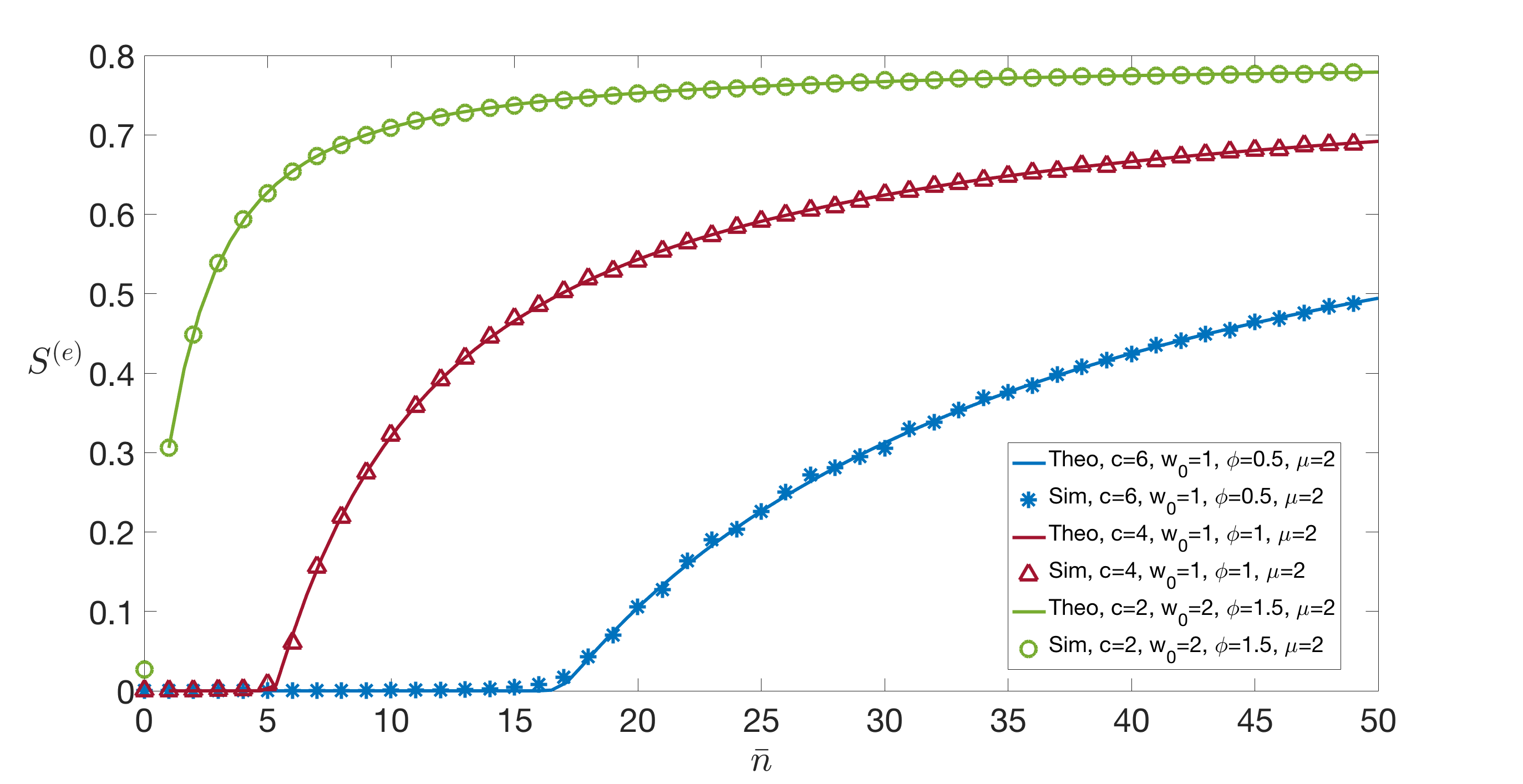}
\caption{Size of the giant component as a function of $\bar{n}$ for different combinations of the parameters $w_0, \phi, c$. Simulations with kernel $f(x,y)$ in Eq. \eqref{definitionf} and exponential wealth distribution with average $w_0$. Numerical results have been obtained for a network of $N=5\cdot 10^4$ nodes averaging over $5$ independent instances. Increasing $w_0$ for similar values of the ratio $c/\phi$ makes the nodes wealthier on average, and therefore more likely to engage in a LN: as a consequence, the average size of the LN-connected component increases for a given value of average volume of transactions. The transition between $S=0$ and $S>0$ happens at a value of $\bar n$ that we denote $\bar n^\star$.}
\label{fig:expsize}
\end{figure}

We now take $\Pi(w)$ to be the exponential pdf with mean $w_0$. The fitness distribution now becomes
\begin{equation}
\rho^{(e)}(x)= \sum_{\ell\geq0} \frac{{\rm e}^{- \bar{n}}\bar{n}^\ell}{\ell!} \int_{0}^{\infty}\frac{{\rm d}w}{w_0} {\rm e}^{-\frac{w}{w_0}}\delta\left(x- w(\ell+1)\right) =\nonumber  \frac{1}{w_0}\sum_{\ell\geq0} \frac{{\rm e}^{- \bar{n}}\bar{n}^\ell}{\ell!(\ell +1)} {\rm e}^{-\frac{x}{w_0(\ell +1)}} \ .
\end{equation}
As in the uniform wealth case
\begin{equation}
\lambda^{(e)}(x) = \frac{1}{f^{(e)}_+}\Theta(x\phi-c)\ ,
\end{equation}
where this time $f^{(e)}_+$ reads
\begin{equation}
f^{(e)}_+= \int_{c/\phi}^\infty {\rm d}x~\rho^{(e)}(x)=
 \frac{1}{w_0}\sum_{\ell\geq 0}\frac{{\rm e}^{-\bar n}\bar n^\ell}{\ell!}\int_0^{\infty}{\rm d}w~{\rm e}^{-w/w_0}\Theta(w(\ell+1)-c/\phi)\ ,\label{Cedef}
\end{equation}
which requires $\ell\geq \lceil \frac{c}{\phi w}-1\rceil$. Therefore,
\begin{equation}
f^{(e)}_+ =1-\frac{1}{w_0}\int_0^\infty\mathrm{d}w~\mathrm{e}^{-w/w_0}\frac{\Gamma\left(\lfloor \frac{c}{\phi w}\rfloor,\bar n\right)}{\Gamma\left(\lfloor \frac{c}{\phi w}\rfloor\right)}\ ,
\end{equation}
where $\lceil z \rceil$ denotes the largest integer smaller than $z$.
As in the uniform-wealth case
\begin{equation}
P^{(e)}(k) =\left(1-f^{(e)}_+\right)\delta_{k,0}+f^{(e)}_+\frac{{\rm e}^{-\mu f^{(e)}_+}(\mu f^{(e)}_+)^k}{k!}\ .\label{Pkexp}
\end{equation}
Now, consider the solution $0<\eta^\star\leq 1$ of
\begin{equation}
\eta^\star = \exp\left[\mu f^{(e)}_+(\eta^\star-1)\right]\ .
\end{equation}
Then, the average size of the giant component reads
\begin{equation}
S^{(e)}=(1-\eta^\star)f^{(e)}_+\ ,
\end{equation}
and the condition for the giant component to appear reads $\mu f^{(e)}_+>1$ (see Fig. \ref{fig:expsize}).

\section{Numerical Simulations and Results} \label{sec:simresults}

We present numerical simulations on networks of $N=5\cdot 10^4$ nodes, generated by sequential deposition of links with probability as in Eq. \eqref{probF}, using the kernel in Eq. \eqref{definitionf}. In Fig. \ref{fig:unifsize}, where we use a uniform distribution of wealth with average $w_0=1$, we plot the average size $S^{(u)}$ of the connected component as a function of $\bar n$, the average volume of transactions to be deployed on the LN, for varying values of the fees ratio $c/\phi$. Fixing a certain average fraction $S^{(u)}$ of nodes -- which can reach each other via a connected LN path -- and increasing the ratio $c/\phi$ between the LN and main-blockchain fees, we observe that a larger average volume of LN transactions is required to make the off-chain network financially sustainable. Increasing the average wealth $w_0$ would push the curves upwards: as more liquidity becomes available across nodes, more and more players may get involved in the LN for the same level of routing fees. In Fig. \ref{fig:expsize}, we observe qualitatively the same phenomenon, this time for exponential distribution of wealth. 

To find the size of the largest connected component, we use a breadth-first search algorithm \cite{search}: starting from a source node $s$, we label it as belonging to cluster $\# 1$. We then explore its neighborhood and assign all nodes reachable from $s$ to cluster $\# 1$ as well. The algorithm proceeds recursively until either the whole network has been labelled, or no unlabelled nodes can be further reached. In the latter case, we select another random source among the unlabelled nodes, assign it the label $\# 2$, and restart the procedure to find another cluster. At the end, all disjoint clusters have been identified, and their size recorded. In our plots, we monitor the size of the largest cluster.

\begin{figure} [h!]
\centering
\includegraphics[width=0.9\textwidth]{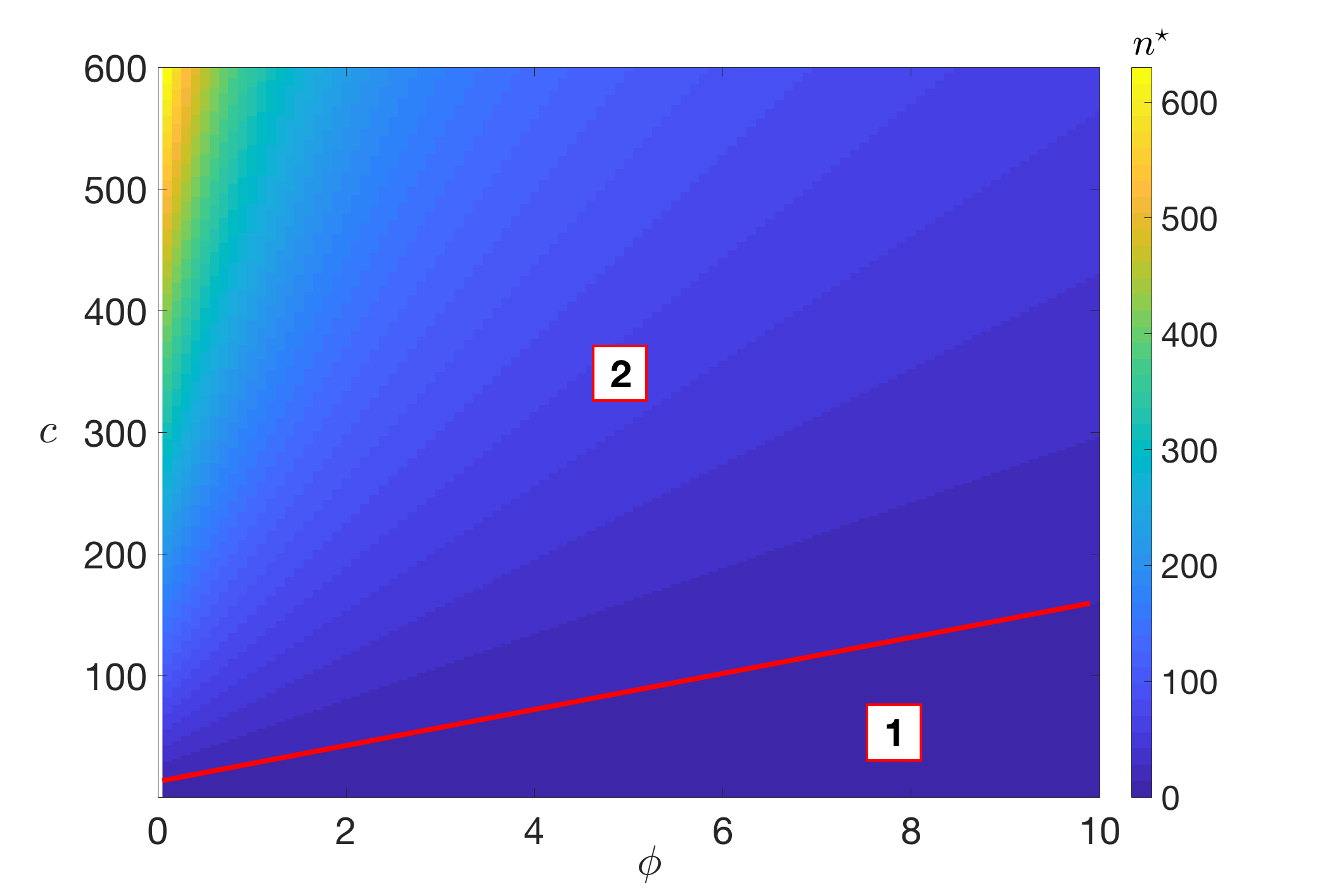}
\caption{Phase diagram in the $(\phi, c)$ plane for uniform wealth distribution in the interval $[0,w_0=1]$, $\mu=20$ and kernel  $f(x,y)$  (Eq. \eqref{definitionf}). The colors from blue to yellow represent (from low to high) the values of the minimal average volume of transactions $\bar{n}^{\star}$ that need to be deployed to make a LN financially viable.
In region $\bf (1)$ (low ratio between LN and main-blockchain fees), $\bar{n}^{\star}=0$, i.e. for any average volume of transactions, a LN connected component is sustainable. In region $\bf (2)$ (larger ratio between Lightning and main blockchain fees), $\bar{n}^{\star}>0$, implying that a LN is not financially viable unless the volume of transactions is sufficiently high. 
}
\label{fig:phaseunif}
\end{figure}

In Fig. \ref{fig:phaseunif}, we plot the phase diagram in the $(\phi,c)$ plane for the uniform wealth distribution model (very similar results are obtained for the exponential wealth distribution, not shown). The colors from blue to yellow represent (from low to high) the values of $\bar{n}^{\star}$, the minimal average volume of transactions that need to be deployed to make a LN financially viable for a given value of LN and main-blockchain fees, $c$ and $\phi$, respectively. We observe a transition between two regimes: one (region {\bf 1}) where the LN fees are sufficiently low (compared to main-blockchain fees) that \emph{any} volume of transactions (however low, $\bar n^\star=0$) can be transferred off-chain and still be financially viable, the other (region {\bf 2}) where the LN fees are sufficiently high that agents may be discouraged from opening channels and  transferring wealth off-chain \emph{unless} there is a minimal volume of transactions to be deployed ($\bar n^\star>0$). The higher the ratio $c/\phi$, the less convenient it is to open LN channels for a fixed value of transactional activity.

\section{Conclusions} \label{sec:conclusions}

In summary, we have presented a simple fitness-based network model for the emergence of a connected set of nodes exchanging wealth off-chain, whose average fractional size $S$ remains finite as $N\to\infty$. The percolation transition resulting from sequential deployment of edges is studied numerically and analytically, as a function of a limited set of parameters that we predict will be in principle possible to infer from empirical or synthetic \cite{LNsimulator} data: $w_0$ (related to the average wealth jointly owned by the agents), $\bar n$ (the average volume of transactions that can be handled off-chain), $c,\phi$ (the fees associated with off-chain and on-chain transactions) and $\mu$ (the average number of channel per node). As a matter of fact, different platforms are currently being offered -- but only at a test stage -- where users can experience the Lighting Network services in a simulated environment. Already at this early stage in the development of a fully operational payment system, some useful data can be gathered: for instance, the platform '1ML'\footnote{\url{https://1ml.com/statistics}} currently aggregates information about $\sim 10 000$ nodes sharing $\sim 30000$ channels, with an average capacity per node of $\sim 1000$ USD, and a base fee per transaction of around $0.000072$ USD. Similarly, for the Bitcoin blockchain we can gather an estimate of $\sim 0.52$ USD as base fee per transaction, as well as more accurate figures about number of transaction per day and average transaction values \footnote{Data available at \url{https://www.blockchain.com/en/charts}.}.

The function $f(x,y)$ in Eq. \eqref{definitionf} has been selected as the simplest but nontrivial attachment kernel that favors a link (i.e. the opening of a Lightning channel) whenever the fitness of both concurring nodes (in terms of exchangeable wealth and volume of predicted activity) exceeds a financially viable threshold. We have checked that ``smoothing" the $0/1$- kernel in Eq. \eqref{definitionf} has negligible effects on the results, while making the analytical treatment unnecessarily more complicated. Similarly, the model is fairly insensitive to the details of the full probability distribution of wealth that is used, while being flexible enough to generate a desired degree distribution $P(k)$ via a different choice of the attachment kernel $f(x,y)$ \cite{smolyarenko}. A percolation transition separates a phase where no sustainable LN can be formed, from a phase where the fees being charged, the total available wealth and the average activity conspire to make off-chain payments a viable option for a finite fraction of the network in the limit $N\to\infty$. The transition is elucidated analytically and numerically, with excellent agreement.

In the future, this investigation can be extended in the following ways:

\begin{itemize}
    \item A mechanism for dynamical update of wealth as more channels are opened and funds are locked may be introduced, to investigate the liquidity constraints of the network in more detail. Dynamically generated wealth inequalities and concentration may be detected by means of centrality measures.
    \item The resilience of the network can be studied under different types of attacks and compared with available empirical results \cite{topology1,topology2}.
    \item Different choices of the kernel $f(x,y)$ (e.g. non-factorized) may be also explored. This could lead to networks with heterogeneous (heavy-tailed) degree distribution, which seems to be in line with recent empirical studies \cite{topology1}.
\end{itemize}
Once the development of the Lighting Network technology and implementation will have reached maturity, it will be possible to gather data to calibrate our model, which can serve as a driver for policy changes and as guidance for incentive mechanisms design. 

\appendix
\section{Degree distribution $P(k)$}  \label{app:degree}
Following \cite{rodgers}, the probability $p_{M,N}(k|x)$ that a node in a large undirected graph with $N$ nodes and $M\ll N^2$ edges has degree $k$ given that its fitness is $x$ follows the recursion 
\begin{equation}
p_{M+1,N}(k+1|x)=p_{M,N}(k+1|x) [1-2\lambda(x,N)] +2p_{M,N}(k|x)\lambda(x,N)\ .
\label{eq:recursion}
\end{equation}
The interpretation is easy: the probability of having a node with degree $k+1$ after an edge addition ($M+1$) is equal to the probability that the node already had degree $k+1$ times the probability that the new edge does not have any of its two terminal points attached to it ($[1-2\lambda(x,N)] $), plus the probability that the node had degree $k$ times the probability that the new edge has one of its two terminal points connected to it ($2\lambda(x,N)$).

Multiplying both sides of Eq. \eqref{eq:recursion} by $s^k$ and summing over $k\geq0$, we obtain the following equation for $F_{M,N}(s|x)=\sum_{k\geq0}s^k p_{M,N}(k|x)$
\begin{align}
\nonumber F_{M+1, N}(s|x)- F_{M, N}(s|x) &= 2F_{M, N}(s|x) (s-1) \lambda(x,N) + F_{M+1, N}(0|x)\\
&- F_{M, N}(0|x) (1-2\lambda(x,N)) \ .
\label{eq:recursion2}
\end{align}
For large $M$, Eq. \eqref{eq:recursion2} can be rewritten as an ordinary differential equation of the form $\frac{\partial F}{\partial M} = 2(s-1)\lambda(x,N) F +(\frac{\partial F}{\partial M} +2\lambda(x,N) F)|_{s=0}$, with solution
\begin{equation}
F_{M, N}(s|x) = {\rm e}^{ \frac{2M}{N}\lambda(x) (s-1)}\ ,
\end{equation}
where we recall that we defined $N\lambda(x)=\lambda(x,N)$ and we used the initial condition $F_{0,N}(s|x)=1$ that follows from the fact that in a network with zero edges, $p_{0,N}(k|x)=\delta_{k,0}$.

Taylor-expanding around $s=0$ and noting that $2M/N = N\kappa$, we obtain the degree distribution conditional on the fitness of the node $x$

\begin{equation}
p_{M,N}(k|x)=\frac{{\rm e}^{-N\kappa\lambda(x)} \left(N\kappa\lambda(x)\right)^k}{k!}\ .
\end{equation}

Marginalizing with respect to $x$, we eventually obtain the probability that a node as degree $k$ (irrespective of its fitness) as
\begin{equation}
P(k) = \int_{0}^{\infty}{\rm d}x \ p_{M,N}(k|x) \rho(x) = \int_{0}^{\infty}{\rm d}x \ \frac{{\rm e}^{-N\kappa\lambda(x)} \left(N\kappa\lambda(x)\right)^k}{k!} \rho(x)\ ,
\end{equation}
which correctly implies
\begin{equation}
\langle k\rangle=\sum_{k\geq 0}k P(k)= N\kappa\ ,
\end{equation}
using \eqref{normlambda}.

\section{Giant component} \label{app:giantc}
The generating function of the probability that a node has degree $k$ is denoted by
\begin{equation}
G_0(s)=\sum_{k\geq 0} P(k) s^k\ .\label{G0s}
\end{equation}
We introduce the generating function $G_1(s)$ of the (normalized) probability that by following a randomly chosen edge we reach a node with degree $k$

\begin{equation}
G_1(s)=\frac{\sum_{k\geq 1}  k P(k) s^{k-1}}{\sum_{k\geq 0} k P(k) } = \frac{G_0^{\prime}(s)}{G_0^\prime(1)} = \frac{G_0^{\prime}(s)}{N\kappa}\ .\label{G1s}
\end{equation}
This is because the node we reach by following a randomly chosen edge has degree distribution $k P(k)/N\kappa$ rather than just $P(k)$ -- since a randomly chosen edge is more likely to lead to a node of higher degree. 

We also define the generating function of the number of nodes that can be reached following a randomly chosen edge and that belong to a connected component of size $t$ with size distribution\footnote{More precisely, 
\begin{equation}
H_1(x)=\lim_{N\to\infty}\sum_{t=1}^N \psi(t,N)x^t\ ,\label{H1finiteN}
\end{equation}
where $\psi(t,N)$ is the probability that -- in a network with $N$ nodes -- by following a randomly chosen link, we reach a component of size $t\leq N$, and similarly for $H_0(x)$ in \eqref{H0x}. } $\psi(t)$ 
\begin{equation}
H_1(x)=\sum_{t\geq 1} {\psi}(t) x^t\ .\label{H1t}
\end{equation}
Moreover, we indicate with $H_0(x)$ the generating function of the probability that a randomly chosen node belongs to a connected component of size $t$
\begin{equation}
    H_0(x)=\sum_{t\geq 1} {\pi}(t) x^t\ .\label{H0x}
\end{equation}
Assuming that the typical component sizes are finite and that the chances of a component containing a closed loop of edges are negligible for sufficiently large $N$, the distribution of components generated by $H_1(x)$ can be obtained as follows \cite{newman,newman2,rodgers}. Let us denote by $\zeta(t|k)$ the probability that a node with degree $k$ belongs to a component of size $t$
\begin{equation}
\zeta(t|k)=\sum_{t_1\geq 1}\cdots \sum_{t_k\geq 1}\delta\left(t-1,\sum_{m=1}^k t_m\right)\prod_{m=1}^k\psi(t_m)\ ,    
\end{equation}
where $\delta(a,b)$ is the Kronecker delta.
Indeed, the sum of the sizes of the components that can be reached by following the $k$ edges departing from the node must be equal to $t-1$, and each of these sizes is drawn from the distribution $\psi(t)$.

Marginalizing over the degree distribution, we obtain the probability $\pi(t)$ that a randomly chosen node belongs to a component of size $t$ as
\begin{equation}
    \pi(t)=\sum_{k\geq 0}P(k)\zeta(t|k)\ .
\end{equation}

Computing $H_0(x)$ from \eqref{H0x}
\begin{align}
\nonumber   & H_0(x) =\sum_{t\geq 1} {\pi}(t) x^t=\sum_{t\geq 1}x^t\sum_{k\geq 0}P(k)\zeta(t|k)\\
 \nonumber  &=\sum_{k\geq 0}P(k)\sum_{t\geq 1}x^t\sum_{t_1\geq 1}\cdots  \sum_{t_k\geq 1}\delta\left(t-1,\sum_{m=1}^k t_m\right)\prod_{m=1}^k\psi(t_m)\\
   &=x\sum_{k\geq 0}P(k)\sum_{t_1\geq 1}\cdots\sum_{t_k\geq 1}x^{\sum_m t_m}  \prod_{m=1}^k\psi(t_m)=x\sum_{k\geq 0}P(k)\left[\sum_{t\geq 1}\psi(t)x^t\right]^k=x G_0(H_1(x))\ ,
\end{align}
where we have used \eqref{G0s} and \eqref{H1t}.
The calculation for $H_1(x)$ is analogous, with the replacement $P(k)\to \frac{k P(k)}{\sum_{k^\prime}k^\prime P(k^\prime)}$. Summarizing, the two equations to be solved together are
\begin{align}
H_0(x) &= x G_0(H_1(x))\ ,\label{eq:ho}\\
H_1(x) &= x G_1(H_1(x))\ .\label{eq:h1}
\end{align}
The average size $\langle t\rangle$ of the connected components is given from \eqref{H0x} as
\begin{equation}
    \langle t\rangle = \sum_{t\geq 1}t\pi(t)=H_0^\prime(1)\ .
\end{equation}
$H_0^{\prime}(1)$ can be obtained from \eqref{eq:ho} as
\begin{equation}
H_0^{\prime}(1) = G_0(H_1(1)) + G^{\prime}_0(H_1(1))H_1^{\prime}(1)\ . \label{H0prime}
\end{equation}
Note that from \eqref{H1t} it follows that $H_1(1)=1$ (by normalization of $\Psi(t)$). Similarly, from \eqref{G0s}, we have that $G_0(1)=1$ (by normalization of $P(k)$). Eq. \eqref{H0prime} can be therefore simplified as follows
\begin{equation}
H_0^{\prime}(1) = 1 + G^{\prime}_0(1)H_1^{\prime}(1)\ . \label{H0primesimple}
\end{equation}
We can then compute $H_1^{\prime}(1)$ using \eqref{eq:h1}
\begin{equation}
H_1^{\prime}(1) = G_1(H_1(1)) + G^{\prime}_1(H_1(1))H_1^{\prime}(1)\ . \label{H1prime}
\end{equation}
As before, we can simplify it using the fact that $H_1(1)=1$ and that $G_1(1)= \frac{\sum_{k}k P(k) s^{k-1}}{\sum_{k}k P(k)}\bigg|_{s=1}=1$ (see\eqref{G1s}), obtaining:
\begin{equation}
H_1^{\prime}(1) = 1 + G^{\prime}_1(1)H_1^{\prime}(1)\Rightarrow H_1^{\prime}(1)=\frac{1}{1-G_1^\prime(1)} \ . \label{H1primesimple}
\end{equation}
Substituting \eqref{H1primesimple} in \eqref{H0primesimple} yields
\begin{equation}
\langle t\rangle=H_0^{\prime}(1) = 1+ \frac{G_0^{\prime}(1)}{1-G_1^\prime (1)}\ , \label{sizeh0}
\end{equation}
which diverges when
\begin{equation}
 1-G_1^\prime (1)=0 
 \label{G1diverge}
\end{equation}
or equivalently (using \eqref{G1s}) when $G_0^{\prime \prime}(1) = N\kappa$, signalling the emergence of the giant component. 

When the giant component has formed, $H_0(x)$ and $H_1(x)$ (see Eq. \eqref{H1finiteN}, \eqref{H1t}, \eqref{H0x}) become the sum of two contributions: one where the sum is restricted to components of size $t\sim o(N)$, and the other restricted to (giant) components of size $t\sim\mathcal{O}(N)$. Assuming that there is only one such giant component, Eq. \eqref{H0x} for $x=1$ can then be written as
\begin{equation}
1 = H_0^{(f)}(1) + S\ ,
\end{equation}
where $H_0^{(f)}(1)$ (and similarly $H_1^{(f)}(1)$) satisfy the equations \eqref{eq:ho} and \eqref{eq:h1}, as they include $\sim o(N)$ contributions for $N\to\infty$ coming from components other than the giant one, whereas $S=N_C/N$ is the fraction of nodes that belong to the giant component. \\
Therefore (from  \eqref{eq:ho} and \eqref{eq:h1})
\begin{equation}
S = 1-G_0(\xi^\star)\ ,
\label{xistarsize}
\end{equation}
where $\xi^\star$ satisfies
\begin{equation}
\xi^\star=G_1(\xi^\star)\ .
\label{xistar}
\end{equation}

\section*{Acknowledgments}
SB and FC acknowledge funding by UCL Centre for Blockchain Technologies as part of the {\em 1st Internal Call for Project Proposals on Distributed Ledger Technologies}. PV acknowledges support from the UKRI Future Leaders Fellowship grant MR/S03174X/1.

\section*{References}
\bibliographystyle{unsrt}
\bibliography{LNpaper-FINAL}

\end{document}